\documentclass[sigconf,nonacm]{acmart}

\setcopyright{none}
\renewcommand\footnotetextcopyrightpermission[1]{}

\acmConference{}{}{}
\acmBooktitle{}
\acmDOI{}
\acmISBN{}
\AtBeginDocument{%
  \providecommand\BibTeX{{%
    \normalfont B\kern-0.5em{\scshape i\kern-0.25em b}\kern-0.8em\TeX}}}

\setcopyright{acmcopyright}
\copyrightyear{}
\acmYear{}
\acmDOI{}

\acmConference[ ]{ }{}{}
\acmBooktitle{} 
\acmISBN{}

\usepackage{subfig,multirow}
\usepackage{xspace}

\newcommand{\f}[1]{{\textsc{#1}}\xspace}

\acmSubmissionID{5411}

\begin{document}

\title[viaCross: Crossing-Based Selection Technique]{viaCross: A Pen-Based Technique for Selecting Objects and Attributes of Interest using Crossing-Based Gestures}


\author{Benedict Leung}
\orcid{0009-0003-6754-9874}
\email{benedict.leung1@ontariotechu.ca}
\affiliation{%
  \institution{Ontario Tech University}
  \city{Oshawa}
  \country{Canada}
}

\author{Mariana Shimabukuro}
\orcid{0000-0003-4104-0781}
\email{mariana.shimabukuro@ontariotechu.ca}
\affiliation{%
  \institution{Ontario Tech University}
  \city{Oshawa}
  \country{Canada}
}

\author{Christopher Collins}
\orcid{0000-0002-4520-7000}
\email{christopher.collins@ontariotechu.ca}
\affiliation{%
  \institution{Ontario Tech University}
  \city{Oshawa}
  \country{Canada}
}

\renewcommand{\shortauthors}{Leung, Shimabukuro, Collins}

\begin{abstract}
% Lasso and marquee tools commonly used in pen-based selection are too general, as every object is selected within the drawn area. In this research, we introduce a pen-based selection technique, viaCross, that uses circle and crossing-based strokes to change spatial and attribute constraints to filter objects from the selection directly during the initial specification and to edit selections. This work also introduces attribute widgets, interface elements that visualize the attribute constraints of a selection and can be used to edit them. A user study that compared viaCross with existing WIMP interfaces showed that viaCross required fewer strokes when targeting objects and more strokes when targeting attributes. Participants also prefer attribute widgets to the panel when targeting attributes, suggesting widgets as a potential replacement for menus/panels in pen-based devices.

Crossing-based selection is well-studied, yet past research has overlooked multi-object and attribute selection with crossing gestures. We present viaCross, a pen-based technique in which crossing strokes define attribute constraints, paired with attribute widgets that visualize and allow subsequent edits to these constraints. A controlled study compared viaCross with traditional WIMP filter panels. Results show that viaCross was more efficient for complex object-level selections, requiring fewer strokes and maintaining efficiency as task difficulty increased. WIMP panels were faster and more intuitive for simple attribute-based selections, but participants encountered repeated deselections and occasional misclicks. viaCross matched panel performance for creating attribute selections but was slower during refinement, as widgets had to be recreated—indicating the need for persistent widgets or hybrid support with a WIMP-style interface. Overall, viaCross demonstrates that crossing-based gestures can provide an efficient and scalable approach to multi-object and attribute selection.

% Our work contributes (1) a novel crossing-based selection technique, (2) an attribute widget mechanism, (3) empirical evidence mapping trade-offs between gestures and panels, and (4) design implications for adaptive interfaces in pen, touch, and immersive contexts.

% \href{https://vialab.github.io/viaCross}{vialab.github.io/viaCross}.
\end{abstract}

%Common user interface techniques employed for selection, such as menus, filter panels, point-and-tap gestures, lasso, marquee tools, and keyboard shortcuts, are not well-suited for pen-based input.

\begin{CCSXML}
<ccs2012>
<concept>
<concept_id>10003120.10003121.10003125.10010873</concept_id>
<concept_desc>Human-centered computing~Pointing devices</concept_desc>
<concept_significance>500</concept_significance>
</concept>
<concept>
<concept_id>10003120.10003121.10003125.10010391</concept_id>
<concept_desc>Human-centered computing~Graphics input devices</concept_desc>
<concept_significance>300</concept_significance>
</concept>
</ccs2012>
\end{CCSXML}

\ccsdesc[500]{Human-centered computing~Pointing devices}
\ccsdesc[300]{Human-centered computing~Graphics input devices}

\keywords{pen-selection techniques, crossing-based gestures, widgets}

\begin{teaserfigure}
  \includegraphics[width=\textwidth]{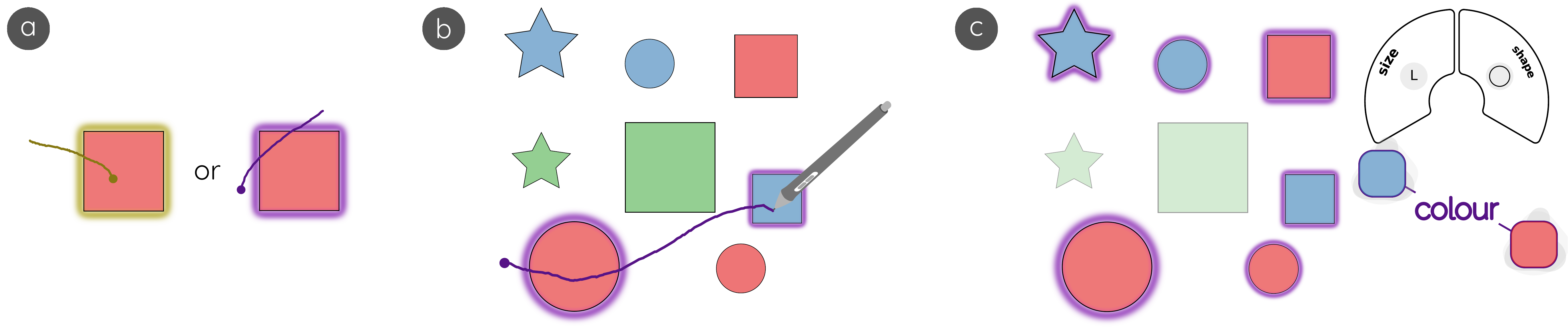}
  \caption{Workflow of using viaCross to select objects in a pen-based environment. a) Specify the selection type by starting the pen on (exact match) or off (attribute similarity) an object. b) Cross objects to extract attributed categories of interest. Here, size, shape, and colour are extracted. c) Specify or edit the constraints of the selection, as a filter. Here, colour is the target, resulting in all blue and red objects being selected.}
  \Description{The figure contains three sub-figures labelled (a), (b), and (c), illustrating the workflow of using viaCross. (a) shows two red squares, the left one outlined in yellow and the right one in purple, illustrating how to initiate (left) attribute similarity and (right) exact match. (b): Displays multiple shapes of three different sizes, shapes and colours. A purple stroke crosses over a large red circle to a small blue square. (c) The same shape layout as (b), but a colour widget is shown, displaying blue and red colours around it, and all blue and red shapes are selected while all green shapes are unselected. }
  \label{fig:teaser}
\end{teaserfigure}


\maketitle

\section{Introduction}

Pen-based devices such as tablets and smart whiteboards have grown in popularity due to their accessibility and improved sensing capabilities. They are widely used for note-taking, sketching, and 3D modelling because pens afford direct, natural interaction with interface elements. Yet, many pen-based applications still rely on WIMP-style menus designed for mouse and keyboard. Without hotkeys or shortcuts, users must perform multiple ``round trips'' between the workspace and menus, increasing interaction costs and disrupting workflow~\cite{PenWithConventionalGUI}.

Crossing-based interaction, selecting a target by crossing its boundary, has been studied theoretically and empirically~\cite{CrossInterfaces, FoundationsCrossingParadigm, SteeringLaw, CrossingDirectTouch, CrossingMotorImpairments, FeedbackCrossing} and applied in interfaces~\cite{CrossY, CrossingToggleMaps, CrossingOverlappignWIndows}. Crossing can perform comparably to point-and-click while enabling more fluid workflows, particularly with pen input. However, little prior work has explored crossing for multi-object selection.

viaCross aims to specify multiple constraints using only one stroke, where the location of the starting point specifies selection type (\autoref{fig:teaser}a), crossing objects specifies attribute constraints (\autoref{fig:teaser}b). As such, users can select multiple attributes using only one stroke. Users employ \textit{attribute widgets} to replace the need for menu round trips while editing attribute constraints (\autoref{fig:teaser}c). Attribute widgets can help declutter the workspace by dividing the traditional menu into necessary and unnecessary options. Further, it allows multiple iteration cycles for a more fluid workflow.

% To evaluate if crossing-based methods could be practical for multi-object selection, we conducted a user study wherein participants used viaCross and a filter panel-based version accompanied by a lasso tool and tapping to select various attributes or objects that match the provided prompt. The results showed that viaCross outperformed the panel at complex object-level selection and refinement, requiring fewer strokes and less time than the panel as task difficulty increased. While panels were faster and more intuitive for simple attribute-based selections, they suffered from repeated deselection and occasional misclicks. viaCross achieved comparable performance for creating attribute selections but was significantly slower during refinement due to the need to recreate widgets, suggesting that preserving attribute widgets at creation or providing a WIMP-style interface post-creation could improve efficiency.
To evaluate viaCross, we conducted a comparative study against a panel-based WIMP interface. Results show that viaCross requires fewer strokes and scales better in complex object-selection tasks, while panels remain faster and more intuitive for simple attribute-based filtering, highlighting complementary strengths.

% Thus, our contribution is viaCross, which is (1) a novel pen-based multi-object, crossing-based selection technique that can be used in a dense environment. It uses (2) attribute widgets that can be used to edit attribute constraints. Finally, (3) we have conducted a study that compares viaCross with traditional WIMP interfaces.

We position our work as both a new workflow that unifies crossing gestures and an exploration of trade-offs in pen-based selection, offering design implications for future interfaces. Thus, our contributions are:  
(1) \textbf{viaCross,  a novel pen-based technique that extends crossing gestures to support multi-object and multi-attribute selection within a single stroke.} % a novel pen-based, crossing-based selection technique for multi-object selection in dense environments.  
(2) \textbf{Attribute widgets} that integrate constraint visualization and refinement directly into the workspace, reducing reliance on WIMP panels. %echanism that enables editing and refining attribute constraints through pen interaction.  
(3) A \textbf{controlled comparative study} with a WIMP-style panel interface, revealing trade-offs: panels are faster and more intuitive for simple attribute filtering, while viaCross scales better in complex, object-centric selections. %with a panel-based WIMP interface, revealing trade-offs between techniques: panels excel in simple attribute filtering, while viaCross scales better for complex, object-centric selections.  
(4) \textbf{Design implications} for hybrid or adaptive interfaces that preserve user effort in refinement and support onboarding for gesture-based systems. %, support for preserving user effort in refinement, and the role of onboarding aids to address learning curves.  

\section{Related Work}

viaCross draws inspiration from prior research on multi-stroke selection techniques and widget design to specify attribute constraints, directly addressing issues from prior crossing-based selection techniques.

\subsection{Crossing-Based Selection}
Crossing-based selection—choosing a target by passing over its boundary—was pioneered by \citet{CrossingFitts}, introducing the idea of ``goal passing.'' This foundation led to extensive experiments across input modalities, including indirect and direct pen~\cite{SteeringLaw, FoundationsCrossingParadigm, FeedbackCrossing, CrossingDialogBoxes}, direct touch~\cite{CrossingDirectTouch}, and mouse~\cite{CrossingToggleMaps, CrossingMotorImpairments}. These studies show that crossing performs comparably to pointing, even in dense environments with distractors where the impact on speed is minimal~\cite{crossingDistractors}. Applications such as CrossY~\cite{CrossY} demonstrate how crossing can be applied to drawing, with all attributes listed in a static panel that users cross through in sequence. However, this approach does not scale well: as the number of attributes grows, panels become larger and spatially constrained, and users are limited to choosing only one attribute per category (e.g., colour, stroke width). 

While prior work demonstrates the promise of crossing, most studies address single-target selection or attribute panels. Our work investigates whether crossing can be extended to multi-object selection and attribute specification, supporting scalability in dense environments.

\subsection{Multi-Stroke Selection Techniques}
Multi-stroke interaction enables richer expression by combining strokes into composite actions. Buxton’s concept of ``phrasing''~\cite{chunking} framed such sequences as cognitive subroutines aligned with a user’s mental model. Building on this idea, \citet{phrasing} introduced techniques like ``selection decoration,'' where one stroke selects an object and another specifies a spatial property. Later work explored mechanisms for distinguishing strokes, ranging from explicit mode switching (e.g., button presses)~\cite{phrasing, modeSwitching}, to recognition-based methods for differentiating writing and gestures~\cite{markingAndDirect, mathPad, withoutPriorMode}. Other strategies leveraged pen landing and liftoff to disambiguate intentions, showing that initial touch location can predict target selection strategies~\cite{renImprovingSelection, renStrategies}. 

viaCross draws from these insights by associating landing strategies with selection type and extending recognition-based approaches to crossing gestures. Unlike prior work focused on multi-stroke gestures, viaCross emphasizes fluid, single-stroke specification where possible, while still supporting refinement through attribute widgets.

\subsection{Menu Access for Pen-Based Input}
Traditional WIMP menus remain costly for pen-based workflows because they require repeated ``round trips'' to static locations~\cite{trackingMenus, ThumbPen, ImprovingMP}. Researchers have explored alternative access methods such as using pen modalities (tilt, roll, pressure)~\cite{TiltMenu, ACoord, penRoll} or gaze-based mode switching~\cite{gazeModeSwitch, gazeShifting}. While promising, these methods face scalability issues as menu items increase or depend on specialized hardware. Gesture-driven widgets provide another solution: Hover Widgets~\cite{hoverwidgets} allowed users to draw gestures to summon local widgets, while stroke-based systems~\cite{PenGestures} supported localized access. However, these techniques required memorization of mappings and suffered from false activations, challenges that grew with the number of functions.

Our \textit{attribute widgets} (\autoref{fig:teaser}) build on this line of research by embedding both information and control into the workspace itself. By coupling constraints with visual feedback, they reduce the need for menu navigation and support more fluid editing of selection filters.

\subsection{Hybrid Interfaces}
Beyond crossing and pen-specific menus, prior work has combined gestural input with menu-based refinement to balance efficiency and control. For example, Kurtenbach and Buxton’s marking menus~\cite{manipulationtechniques} blend stroke gestures with radial menus to support both novice and expert use. Hover Widgets~\cite{hoverwidgets} and Scriboli~\cite{PenGestures} similarly integrated pen gestures with localized or delimiter-based controls to reduce round trips to static panels. More recently, Attribute Spaces in VR~\cite{attributeSpaces} provided in-situ manipulation tools for refining selections within immersive environments. These hybrid approaches illustrate a recurring design strategy: combining the fluidity of gestures with the explicitness of structured controls.

viaCross follows this trajectory by unifying crossing-based multi-object selection with in-place widgets for attribute refinement. In doing so, it extends prior hybrid approaches to pen-based environments, while our evaluation highlights how trade-offs between panels and gestures motivate further exploration of adaptive or blended interfaces.

\section{viaCross: Crossing-Based Selection Technique}\label{sec:techniques}
\begin{figure*}[ht]
    \includegraphics[width=0.97\textwidth]{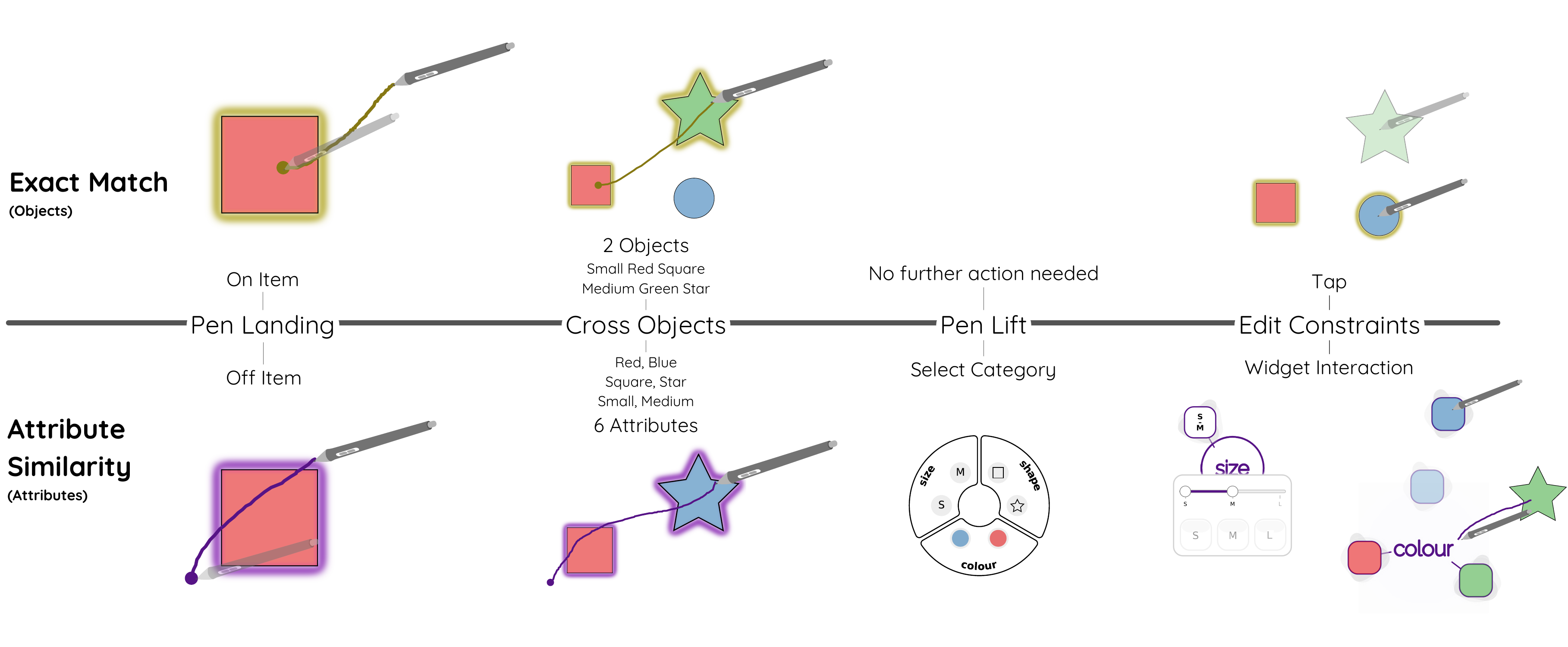}
    \caption{Design space of viaCross showing techniques and gestures for targeting attributes of interest. Widget interactions differ for ordinal/quantitative attributes (sliders available), and nominal attributes (no sliders).
    % \textit{Spatial} refers to which gestures should be performed to change the scope of the selection.
    }
    \Description{
        The figure is divided into two horizontal rows, each depicting a complete four-stage interaction sequence in viaCross: pen landing, crossing objects, pen lift, and edit constraints.
        
        Top row - Exact Match (Objects): The sequence begins with the pen landing directly on a red square. The pen stroke may cross one or more objects. Lifting the pen confirms the selection. Tapping on other objects removes or adds the constraints from the selection.
        
        Bottom row - Attribute Similarity (Attributes): The same four-stage flow is applied, but initial selection is based on attribute match: the pen lands on the background and crosses over a red square. Crossing passes over other objects to extract the attribute set (e.g., colour, size, shape). Pen lift finalizes the selection through the radial menu, and the widget interaction adjusts constraints with the same slider/no-slider distinction based on attribute type.
    }
    \label{fig:desginspace}
\end{figure*}

This section outlines the workflow of viaCross, which consists of three steps, as shown in \autoref{fig:desginspace}: 1) Specify selection type by indicating pen landing. 2) Specify selection constraints by crossing or picking categories. 3) Optionally, edit post-selection constraints. viaCross integrates pen stroke interactions with an implicit attribute or object selection via contextual widget menus, minimizing round-trips. It interacts directly with workspace objects and modifies attribute constraints based exclusively on pen strokes, without a keyboard or additional gestures. Demo available at \href{https://vialab.github.io/viaCross}{vialab.github.io/viaCross}.

\subsection{Attribute Constraint Selection}
Attribute constraints are used to target attributes or objects of interest, and viaCross supports three types. Nominal constraints deal with unordered values (e.g., colours, shapes), ordinal constraints with ordered values (e.g., sizes), and quantitative constraints with numerical values, either discrete or continuous (e.g., ratings). Interactions vary by attribute type. Specifying constraints with viaCross involves three steps (\autoref{fig:desginspace}). First, the user selects the desired selection type. Next, they add constraints. We aim to allow initial selection type and constraint specification in one gesture. Lastly, the system visualizes ambiguous attribute similarity constraints for user editing.

\textit{\textbf{Selection Type Creation}}
%Different selection types can be used to interpret constraints differently, for example, how similar objects are to the constraints or to determine if any duplicates match the constraints. 
viaCross offers two selection types: \textit{exact match} and \textit{attribute similarity}. The selection type depends on the pen stroke's initial coordinate (\autoref{fig:desginspace}, pen landing). If it intersects an object, an exact match selection is created. Otherwise, an attribute similarity selection is used. The intuition of the design choice is that starting on an object specifies a direct intention for that object exactly, while passing through suggests a broader interest in its attributes. After identifying the selection type, constraints can be added by extending the stroke to intersect other objects. Selected objects will be finalized once the pen is lifted. Using the pen's eraser tip on the boundary or selected items will delete the selection.

\begin{figure*}
    \centering
    % \subfloat[Initiating attribute similarity selection]{\includegraphics[width=0.32\textwidth]{images/attributePieMenu.png}\label{fig:attributePieMenu}}
    % \subfloat[Colour selection via a pie menu]{\includegraphics[width=0.32\textwidth]{images/attributePieMenu2.png}\label{fig:attributePieMenu2}}
    \includegraphics[width=0.97\textwidth]{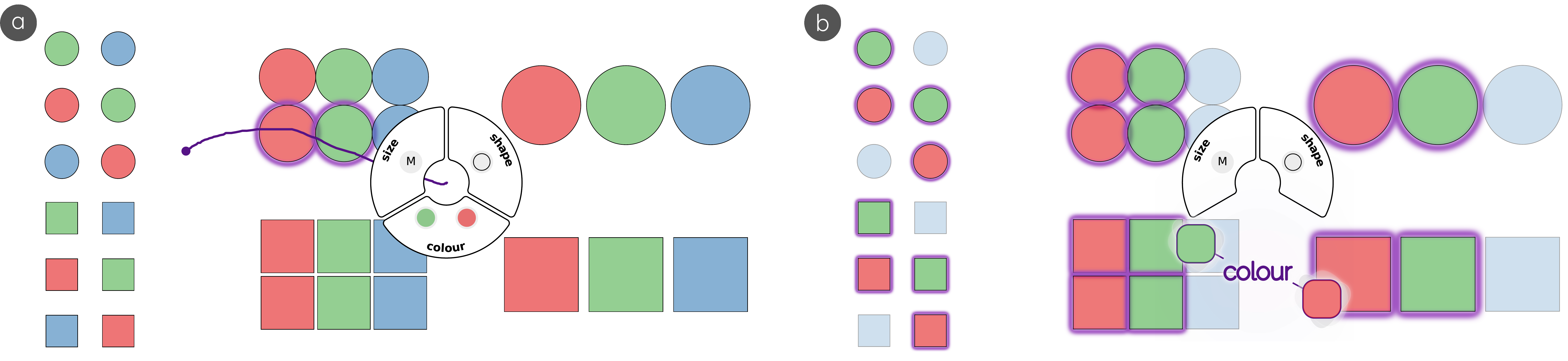}
    \caption{Example of attribute similarity selection. (a) A crossing stroke intersects a medium red and a medium green circle, adding colours, shape, and size attributes to the attribute pool. (b) Selecting colour from the pie menu targets all items matching red or green, regardless of other attributes.}
    % \caption{Once an attribute similarity selection initiates a pie menu, displaying all categories. After selecting a category, an interactive widget visualizes the selection constraints. (a) For instance, an attribute similarity stroke intersects a medium red and green circle, adding two colours, one shape, and one size to the attribute pool. (b) Selecting colour from the pie menu selects all colours from the pool, ignoring other attributes and selecting all green and red shapes.}
    \Description{Two subfigures, labelled (a) and (b), depict an example of attribute-similarity selection in a grid of colored shapes. In subfigure (a), a crossing stroke runs through a medium red circle and a medium green circle. A radial pie menu appears at the stroke’s end, with segments labelled for attributes such as colour, shape, and size. In subfigure (b), the colour segment is shown as selected, and a colour widget is created, and all shapes matching either red or green are highlighted across the grid, regardless of shape or size.}
    \label{fig:att-similarity}
\end{figure*}

\textit{\textbf{Exact Match Creation}}. Exact match selection finds objects where all attributes match those of the target (e.g., all `small green circles'). Users can intersect multiple objects to expand their selection by extending their pen strokes. Once the pen is lifted, objects matching any crossed items are selected. Unlike traditional menus requiring all attributes for each item, viaCross allows users to intersect all items with a single stroke. \textit{\textbf{Editing Exact Match}}. Users can tap objects to add or remove constraints to modify an exact match selection.

\textit{\textbf{Attribute Similarity Creation}}. Attribute similarity matches objects with similar attributes, such as all `green' shapes (colour). Users can define attributes by intersecting objects with pen strokes (\autoref{fig:att-similarity}), but this can lead to ambiguous specifications. For instance, if the stroke passes through a medium red circle and a medium green circle, the user might want green and red objects (colour), medium objects (size), or circles (shape). After lifting the pen, a pie menu shows all data categories (e.g., shape, size, colour; \hyperref[fig:att-similarity]{\autoref*{fig:att-similarity}a}). Users can select target attributes, and a widget appears upon selection (\hyperref[fig:att-similarity]{\autoref*{fig:att-similarity}b}). Unlike traditional methods that require searching for each attribute in a list, viaCross can extract relevant attributes in one stroke. Although the example shows three categories, improved pie menu designs can accommodate more. 
\textit{\textbf{Editing Attribute Similarity}}. Widgets can also be used to edit attribute constraints. More details are described in \autoref{sec:widgets}.

%, shown in \nameref{fig:attributePieMenu2}, and is further described in \autoref{sec:widgets}. By default, widgets will select attributes presented in intersected objects. For example, in \nameref{fig:attributePieMenu2}, the pen stroke has intersected two colours, one shape, and one size. Creating a colour widget will select those two colours, green and red, making the final selection to select all green and red shapes and ignoring size and shape. The pie menu will stay until the user makes another stroke on the screen.

\begin{figure*}
    \centering
    % \subfloat[Colour widget]{\includegraphics[width=0.168\textwidth]{images/widget.png}\label{fig:colourwidget}}
    % \hspace{0.5cm}
    % \subfloat[Colour widget menu]{\includegraphics[width=0.25\textwidth]{images/widgetMenu.png}\label{fig:widgetmenu}}
    % \hspace{0.5cm}
    % \subfloat[Add attribute via stroke]{\includegraphics[width=0.23\textwidth]{images/widgetDraw.png}\label{fig:widgetdraw}}
    % \hspace{0.5cm}
    % \subfloat[Remove attribute via drag]{\includegraphics[width=0.15\textwidth]{images/widgetDrag.png}\label{fig:widgetdrag}}
    \includegraphics[width=0.97\textwidth]{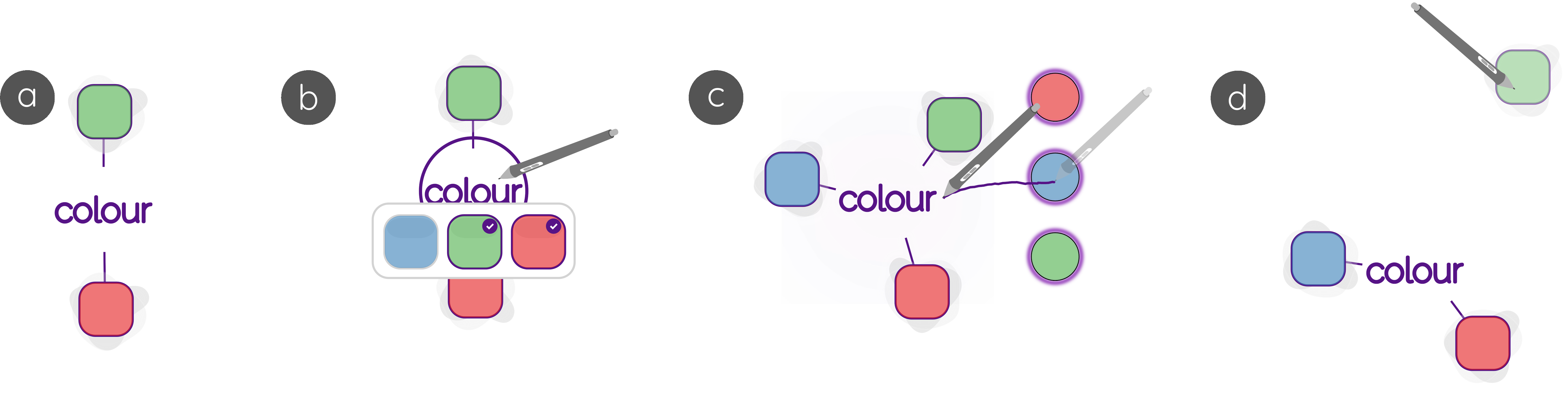}
    \caption{Editing attribute widgets in viaCross. (a) Initial colour widget with red and green selected. (b) Hover or tap on the widget to open the menu of available attributes. Alternatively, to edit a widget, you can either (c) draw a stroke from the object to the widget and add attributes. To remove attributes, (d) drag icons away from the widget.}
   \Description{Four subfigures, labelled (a) through (d), show stages of interaction with a colour attribute widget in viaCross. (a) The initial state displays a small widget containing red and green swatches, indicating the currently selected colours. (b) The widget is displayed with a menu of available attribute icons, which are revealed by a hover or tap gesture. (c) A stroke is drawn from a shape in the workspace toward the widget, and an additional colour swatch appears inside the widget. (d) The green swatch is dragged away from the widget area, removing it from the selection constraints.}
    \label{fig:widgetdemo}
\end{figure*}

\begin{figure}
    \centering
    % \subfloat[Pie menu via holding down the pen]{\includegraphics[width=0.3\columnwidth]{images/afterPieMenu.png}\label{fig:afterpiemenu}}
    % \subfloat[Widget creation via a stroke]{\includegraphics[width=0.2\columnwidth]{images/widgetStroke.png}\label{fig:widgetstroke}}
    % \subfloat[Result of stroking to create a widget]{\includegraphics[width=0.45\columnwidth]{images/sizeWidget.png}\label{fig:sizewidget}}
    \includegraphics[width=\columnwidth]{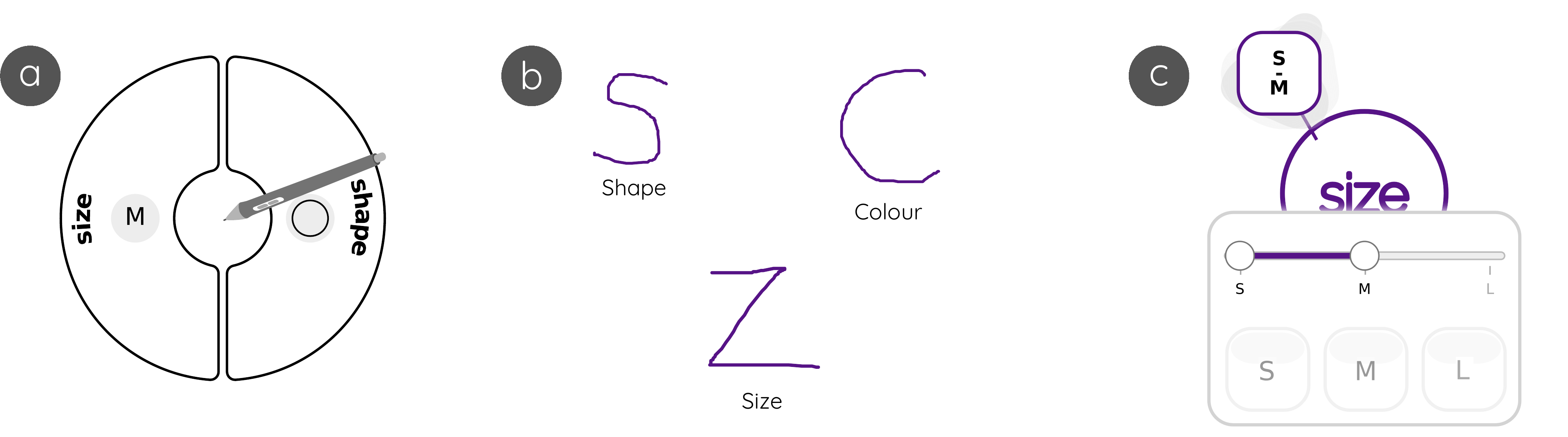}
    \caption{Methods for creating attribute widgets in viaCross: (a) A pie menu shows non-targeted categories. Tapping a category creates the associated widget. (b) Alternatively, draw the letter for its attribute (e.g., `C' for colour, `S' for shape, `Z' for size), which creates a size widget when `Z' is drawn. (c) The resulting widget for quantitative/ordinal attributes includes a slider for range selection.}
    \Description{Three subfigures, labelled (a) through (c), illustrate methods for creating attribute widgets in viaCross. (a) A circular pie menu is divided into two labelled segments: size and shape, indicating non-targeted categories. (b) Examples of handwritten letters ``S'', ``C'', and ``Z'' appear on the workspace, each corresponding to a category. (c) A size widget for quantitative or ordinal attributes is shown, consisting of a horizontal slider ranging from small to large above a series of size icons.}
    \label{fig:widgetdemo2}
\end{figure}

\subsection{Widget Visualization}\label{sec:widgets}
viaCross uses widgets to edit and visualize attribute similarity constraints (\autoref{fig:widgetdemo}). Each widget is responsible for only one attribute category (e.g., colour, shape) (\hyperref[fig:widgetdemo]{\autoref*{fig:widgetdemo}a}). It features a label for the category name, icons for selected attributes, and a hidden menu for novice users (\hyperref[fig:widgetdemo]{\autoref*{fig:widgetdemo}b}). Users can move widgets by dragging their labels. Users can interact with the widget and the application simultaneously to add or remove attributes as they near the user's active space, unlike static menu bars that appear at the top or side of an application.

Users can access the hidden menu or perform a gesture to edit the filter on the widget. The menu appears when the widget is hovered over or its label is tapped. It shows which attributes are selected or unselected, allowing users to add or remove them by tapping icons. Attributes can also be added by drawing a stroke from the object to the widget (\hyperref[fig:widgetdemo]{\autoref*{fig:widgetdemo}c}). Users can drag icons away to remove attributes until the link is broken (\hyperref[fig:widgetdemo]{\autoref*{fig:widgetdemo}d}). Once done, users can erase the widget by dragging the pen's eraser across its label without altering the filters. 

The initial pie menu will disappear with any pen stroke drawn on the screen. To allow new widget creation post-disappearance, viaCross offers two methods (\hyperref[fig:widgetdemo2]{\autoref*{fig:widgetdemo2}}). Novice users can activate the pie menu by briefly holding the pen tip against the screen (\hyperref[fig:widgetdemo2]{\autoref*{fig:widgetdemo2}a}), showing unselected categories. In contrast, expert users can draw a letter related to the attribute category—‘C' for colour, `S' for shape, and `Z' for size (\hyperref[fig:widgetdemo2]{\autoref*{fig:widgetdemo2}b}). In this case, the stroke will morph into a widget once the letter has been recognized. A slider appears for quantitative and ordered attributes to select ranges (\hyperref[fig:widgetdemo2]{\autoref*{fig:widgetdemo2}c}).

\section{User Study Design}

\begin{figure}
    \centering
    \includegraphics[width=0.18\textwidth]{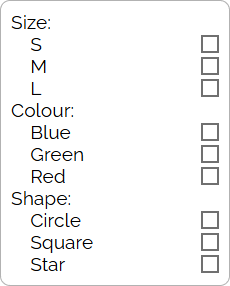}
    \caption{Panel-based baseline interface used in the study. A fixed filtering panel in the top-right corner lists all attributes; participants could tap or lasso items to update selections.}
    \Description{The panel is divided into three labelled sections: size, colour, and shape. Under size is available in three options: S, M, and L. Under colour are Blue, Green, and Red. Under shape are circle, square, and star. Each option is accompanied by a small checkbox.}
    \label{fig:panel}
\end{figure}

To evaluate viaCross, we compared it to a traditional WIMP interface. The panel-based technique used a pen-based lasso tool and a filter panel, inspired by existing query interfaces~\cite{visualExploration, phlat, queryPreview, dynamicFilters}, where all constraints are listed (Figure \ref{fig:panel}). The panel, located in the top right corner of the screen, contained filters applied based on the participant's selection. Participants could also tap on individual objects to add or remove them from the selection. The main comparison variables are the number of strokes and the task duration. Our study was approved by our institutional Research Ethics Board (File \#16971).

% \subsection{Hypothesis}
% Our study compares this panel technique with viaCross to determine which method requires fewer interactions across various tasks. We measure this by comparing task duration and the number of strokes performed to complete the task for each method. Thus, the experiment investigated two hypotheses:

% \paragraph{\textbf{Number of strokes}} viaCross will result in fewer strokes when targeting objects. When targeting attributes, viaCross will result in more strokes only for easier tasks due to the activation step. Attribute similarity needs one crossing stroke and a tap on the pie menu, while the panel only needs one tap per attribute. 

% \paragraph{\textbf{Task duration}} The panel will result in a shorter task duration due to the familiarity of the technique.

\subsection{Participants}
24 participants were recruited to participate in the study via an advertisement on a public poster at a local university campus. 22 participants were right-handed and two were left-handed. 5 participants used a digital pen daily, 3 used weekly, 9 used monthly, and 7 had never used them. A \$30 CAD honorarium was awarded for participation.

% \begin{figure}[b]
%     \centering
%     \subfloat[]{\includegraphics[height=5cm]{images/surface.jpg}\label{fig:surface}}
%     \hspace{1cm}
%     \subfloat[]{\includegraphics[height=5cm]{images/room.jpg}\label{fig:room}}
%     \caption{The \protect\subref{fig:surface} Microsoft Surface Pro 6 and Surface Pen (1st generation) and \protect\subref{fig:room} Room Setup used during the experiment. The keyboard was used to answer questionnaires.}
%     \label{fig:setup}
% \end{figure}

\subsection{Apparatus and Software}
Participants were seated at a desk using a Microsoft Surface Pro 11 and Surface Pen throughout the experiment, and these devices were used to run the study system.

During the main study tasks, the system randomly placed 10 or 11 objects (quantity also random). To ensure distribution consistency, the layout has at least three objects with identical attributes across all categories (e.g., three green, three circles, three large) in each quadrant. To minimize shape overlapping, 8000 iterations of Boids's~\cite{boids} algorithm were used to refine the layout, where the model ensured that each shape avoided others within an awareness radius, did not cross the region dividing boundaries or the screen boundary, and each shape was attracted to the center of the screen.

\subsection{Experimental Design}
% The study used a repeated-measures design, wherein each participant performed 4 trials of the creation selection task per \f{selection technique} $\times$ \f{selection type} $\times$ \f{selection scope} and 2 trials of the refine selection task per \f{selection technique} $\times$ \f{selection type} $\times$ \f{selection scope}. 
% Originally 4 trials were used per \emph{selection technique} $\times$ \emph{selection type} $\times$ \emph{selection scope}, but we found that the study took 120 minutes during the pilot, which increases fatigue effects and as a result reduced the refine selection task. 
We used a within-subject design with two primary independent variables: \f{technique} with two levels (viaCross, panel) and \f{task} with two levels (create, refine). The presentation order of the \f{technique} was counter-balanced to minimize fatigue and learning effects.

\f{type} and \f{difficulty} formed secondary independent variables. There were two levels of \f{type} (attributes, objects). For \f{difficulty}, 1/2/3 attributes or objects were used for the create and refine tasks. The difficulty was gradually increased from easiest to hardest, helping users learn the technique for more challenging tasks by first completing the easier ones. This scaling mirrors real-world selection scenarios, where increased object density and attribute overlap make filtering tasks progressively harder.

The primary measures computed from logs are \emph{overall task duration} and \emph{overall number of strokes} that were made on each trial. A stroke was counted as every tap, lasso, drag, and line drawn. To focus on the final submitted answer, we truncated each trial to analyze the number of strokes and task duration that occurred after the last erasing action, resulting in two additional measures (\emph{post-erasing strokes}) and (\emph{post-erasing duration}). These measures focus on the final, post-correction attempt for each trial. The data was then split across \f{type}. ART ANOVA and its post-hoc tests were then conducted with \f{technique} and \f{difficulty} with Holm-Bonferroni adjustment. We also computed each trial's \emph{accuracy} by comparing the submitted selection against the prompted selection to determine if they matched, resulting in a binary value (i.e., yes or no).

In summary, 2 trials per \f{technique} $\times$ \f{task} $\times$ \f{type} $\times$ \f{difficulty} for each participant. In total, 48 trials $\times$ 24 participants = 1152 trials.

% During the study, the system recorded every stroke that participants made, wherein a stroke was counted as every tap, lasso, drag, and line drawn. For each recorded stroke, the type of stroke was classified based on what gesture was recognized and performed (e.g., ``create exact match'', ``create a widget via pie menu''). We also recorded the time to start and finish executing the stroke and coordinates. From this data, we measured the \emph{overall task duration} and \emph{overall number of strokes} that were made on each trial. As participants could restart a selection by erasing it during the create selection task, we also truncated each trial and analyzed the number of strokes and task duration that occurred after the last erasing action, resulting in two additional variables (\emph{post-erasing strokes}) and (\emph{post-erasing duration}). We also computed each trial's \emph{accuracy} by comparing the submitted selection against the prompted selection to determine if they matched, resulting in a binary value (i.e., yes or no). One trial from each of three participants was not recorded due to a technical error, so these trials were excluded from the dataset.

% The data was then split across \f{selection type}: (attributes, objects). A 3-way RM-ANOVA was then conducted with \emph{selection technique: (viaCross, panel)}, \emph{selection scope: (global, local)}, and \emph{difficulty: (1--4 attributes; 1--3 objects; 1--2 refine attributes/objects)} to evaluate our hypotheses.

\begin{figure}[t]
    \centering
    %\subfloat[]{\includegraphics[width=0.6\textwidth]{images/traininglayout.png}\label{fig:trainlayout}}
    %\\
    \includegraphics[width=.97\columnwidth]{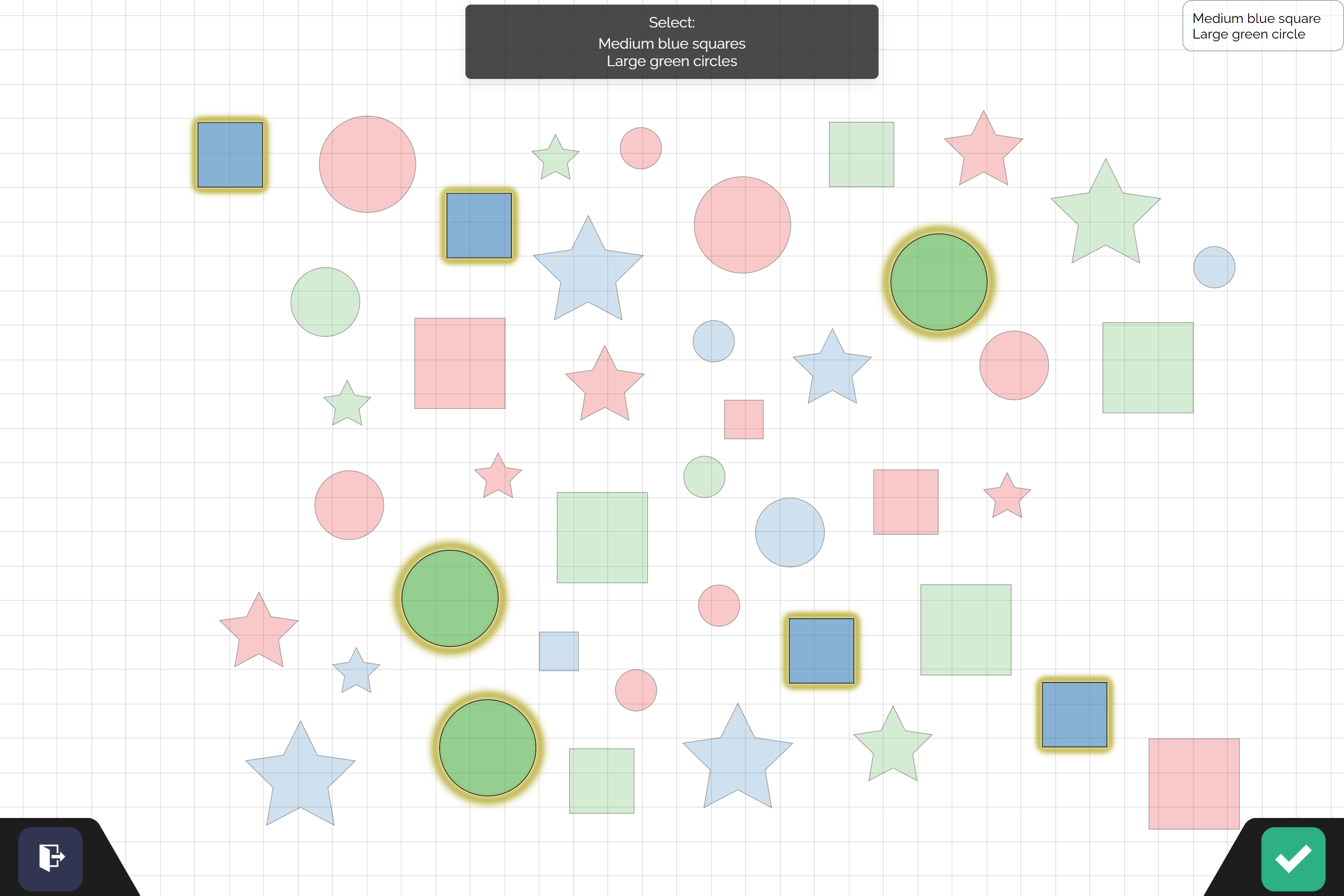}\label{fig:testlayout}
    \caption{Example of a generated trial layout showing objects distributed across the workspace. Each quadrant contains at least three objects sharing identical attributes in each category to ensure balanced conditions.}
    \Description{A grid-based workspace is populated with stars, squares, and circles in red, green, and blue, each rendered in small, medium or large sizes. The shapes are distributed evenly into four quadrants. At the top center, a selection menu lists two criteria: "Medium blue squares" and "Large green circles." Shapes meeting either criterion are outlined with a yellow border.}
    \label{fig:layout}
\end{figure}

\subsection{Procedure}
\paragraph{Introduction} After completing the demographics questionnaire, the participants followed the system instructions.
% read on-screen instructions on how the techniques work.

\paragraph{Training} Next, participants were asked to match a prompt displayed on the screen. Feedback was provided during training as to whether tasks were completed correctly. The training was split into two blocks:
\begin{enumerate}
    \item \textit{6 trials}: create and edit an exact match selection
    \item \textit{6 trials}: create attribute similarity selection, create attribute widgets, and edit filters
    % \item \textit{3 trials}: edit spatial constraints of an existing selection. For viaCross, use circling and bump and bite gestures. Use lasso, tapping, and a filter panel for the panel technique.
\end{enumerate}

The tasks and target layouts remained the same throughout the practice periods. Participants could practice the techniques after each training block, and all practice prompts were the same for both techniques.

\paragraph{Evaluation} Participants began trials using one of two selection techniques. Each trial displayed a window containing the task prompt and a dismiss button. After dismissing, participants viewed a collection of objects organized by the layout procedure, with the prompt at the top. When ready, they pressed a button to finalize their selection and proceed to the next trial.

Participants completed the creation and then the refinement tasks. In creation, they selected objects matching descriptions from all attribute categories (e.g., small red circles and large blue stars (2 objects)) or attributes (e.g., green squares (1 colour and 1 shape)). Despite variations in selection types, participants determined whether to use exact match or attribute similarity for viaCross. In refinement, they adjusted a randomly generated selection to align with the prompt, ensuring it contained the same number of attributes or objects as the selection.

\paragraph{Questionnaires} After each technique, they were asked to rate their overall experience on a 5-point Likert scale. After both techniques, participants completed a questionnaire about their preferences for the techniques. The experimental session lasted approximately 60 minutes.

\section{User Study Results}
\begin{figure}[t]
    \centering
    \includegraphics[width=\columnwidth]{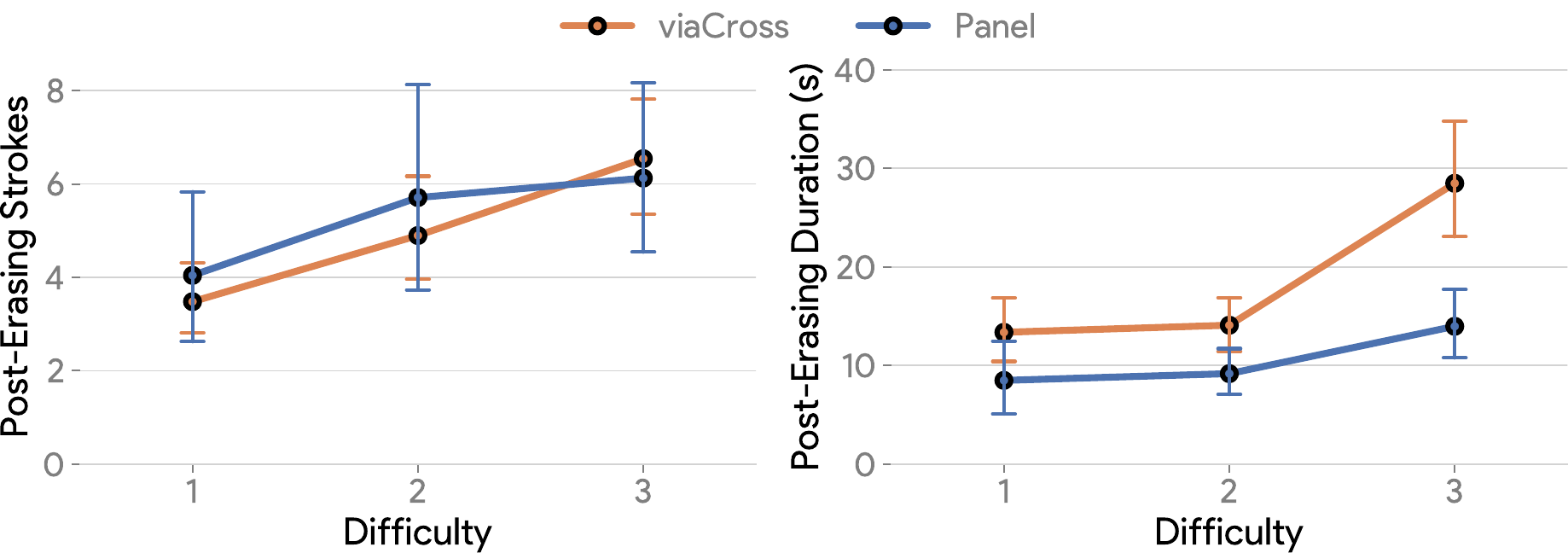}
    \caption{Plot relating to targeting attributes focused on (left) the number of post-erasing strokes and (right) the post-erasing duration for each difficulty. Error bars depict the 95\% confidence interval. viaCross and panel show similar stroke counts across difficulties, with only minor practical differences ($\approx$0.3 strokes). 
Duration differences are small, indicating comparable efficiency for attribute-level creation.}
    \Description{Two side-by-side line graphs compare viaCross and panel across three difficulty levels (1--3). The left graph plots the number of post-erasing strokes (y-axis, 0--6) against difficulty level (x-axis). Both methods show an upward trend as difficulty increases. The right graph plots post-erasing duration in seconds (y-axis, 0--40) against difficulty level (x-axis). Both methods similarly rise with difficulty, with the viaCross line increasing more steeply at difficulty 3.}
    \label{fig:createAttributePlotsPost}
\end{figure}

\begin{figure}[t]
    \centering
    \includegraphics[width=\columnwidth]{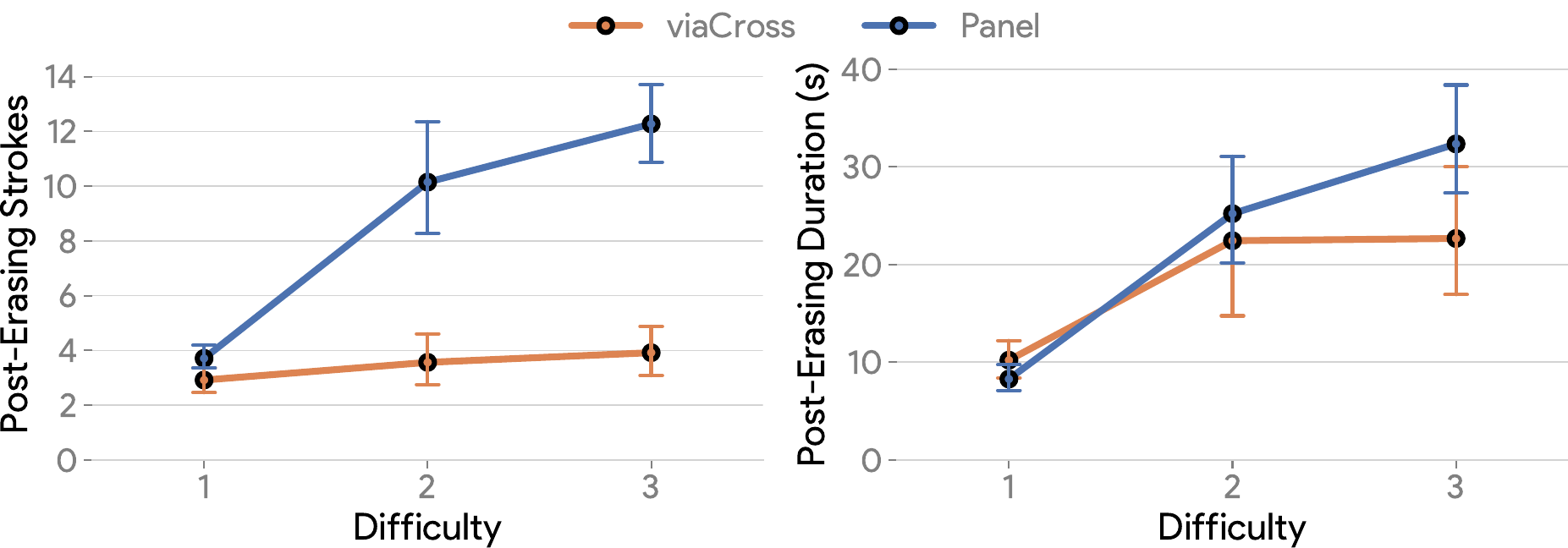}
    \caption{Plot relating to targeting objects focused on (left) the number of post-erasing strokes and (right) the post-erasing duration for each difficulty. Error bars depict the 95\% confidence interval. viaCross consistently requires substantially fewer strokes than the panel, with the gap widening as difficulty increases (up to $\approx$8 strokes saved at the hardest level). Duration patterns follow a similar trend.}
    \Description{Two side-by-side line graphs compare viaCross and panel across three difficulty levels (1-3). The left graph plots the number of post-erasing strokes (y-axis, 0-14) against difficulty level (x-axis). Both methods show an upward trend as difficulty increases, but the panel is significantly steeper than viaCross. The right graph plots post-erasing duration in seconds (y-axis, 0-40) against difficulty level (x-axis). Both methods similarly rise with difficulty, but viaCross flattens at difficulty 3.}
    \label{fig:createExactPlotsPost}
\end{figure}

\begin{figure}[t]
    \centering
    \includegraphics[width=\columnwidth]{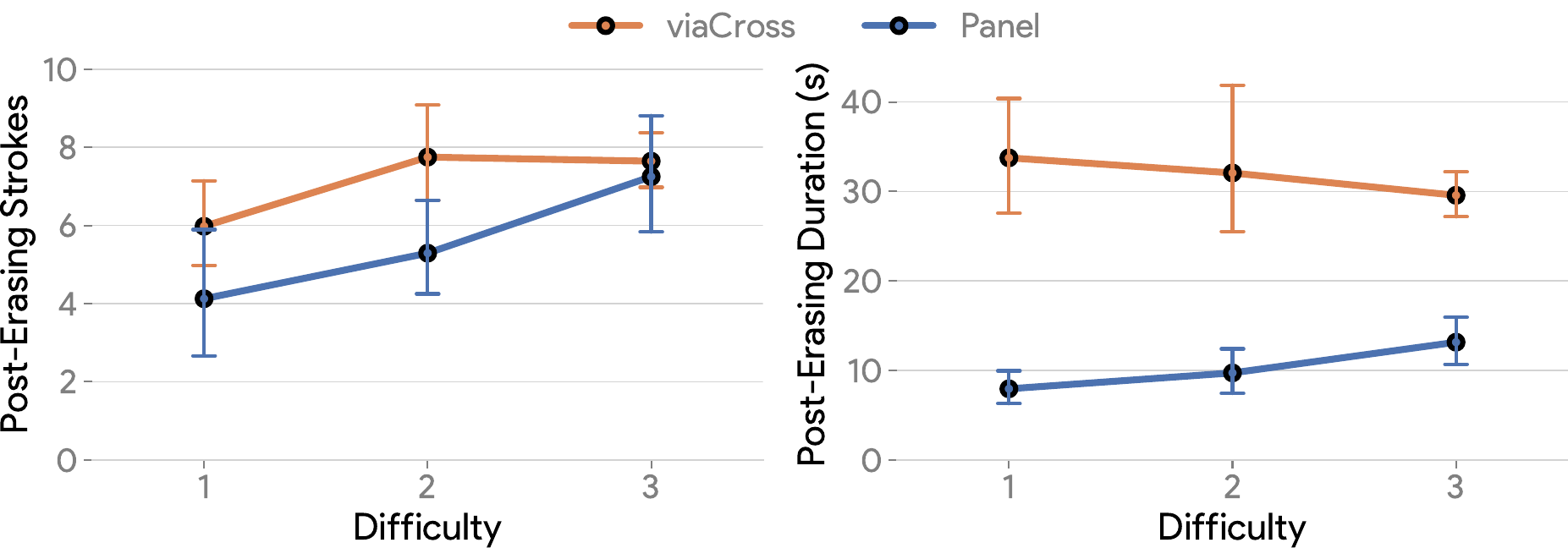}
    \caption{Plot relating to refining attributes focused on (left) the number of post-erasing strokes and (right) the post-erasing duration for each difficulty. Error bars depict the 95\% confidence interval. Panel requires fewer strokes overall than viaCross for refinement of attributes. Duration patterns suggest a similar advantage for the panel in this task type.}
    \Description{Two side-by-side line graphs compare viaCross and panel across three difficulty levels (1-3). The left graph plots the number of post-erasing strokes (y-axis, 0-10) against difficulty level (x-axis). Both methods show an upward trend as difficulty increases, but viaCross shows a higher trend and closes the gap towards difficulty 3. The right graph plots post-erasing duration in seconds (y-axis, 0-40) against difficulty level (x-axis). ViaCross exhibits a downward trend, whereas the panel displays an upward trend. However, ViaCross is significantly higher than the panel.}
    \label{fig:refineAttributePlots}
\end{figure}

\begin{figure}[t]
    \centering
    \includegraphics[width=\columnwidth]{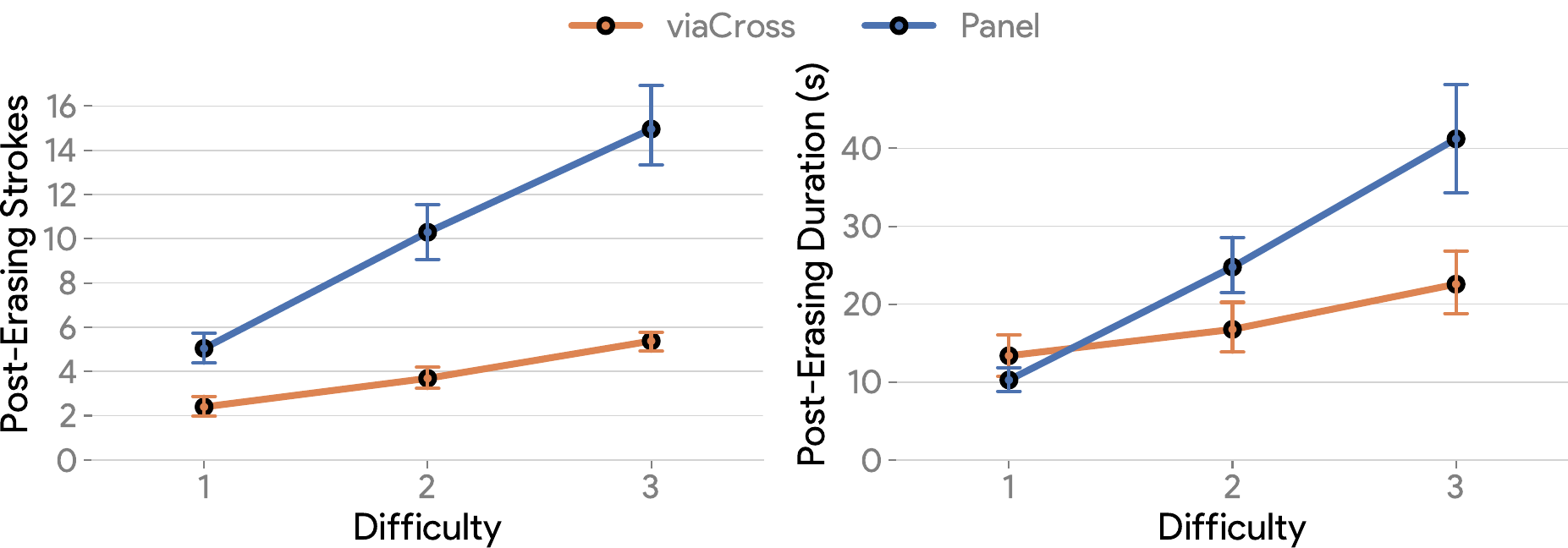}
    \caption{Plot relating to refining objects focused on (left) the number of post-erasing strokes and (right) the post-erasing duration for each difficulty. Error bars depict the 95\% confidence interval. viaCross requires far fewer strokes than the panel across all difficulties (saving 5--10 strokes), with the greatest relative benefit at higher difficulty levels. Duration patterns suggest the stroke advantage may not always translate to time savings.}
    \Description{Two side-by-side line graphs compare viaCross and panel across three difficulty levels (1-3). The left graph plots the number of post-erasing strokes (y-axis, 0-16) against difficulty level (x-axis). Both methods show an upward trend as difficulty increases, but the panel shows a steeper trend. The right graph plots post-erasing duration in seconds (y-axis, 0-40) against difficulty level (x-axis). Both methods show an upward trend, but the panel is shows a steeper trend.}
    \label{fig:refineExactPlots}
\end{figure}

In summary, the results show that viaCross saves strokes in complex object tasks but costs time in attribute refinement. The results are organized by the measures and then by an accuracy analysis and the 5-point Likert scale responses. The measures are summarized in Figures \ref{fig:createAttributePlotsPost}--\ref{fig:refineExactPlots}.

\subsection{Number of Strokes}
When targeting attributes, there was a significant main effect of \f{technique} ($F(1, 259) = 40.06; p < .0001; \eta^2_p = .13$). Post-hoc tests showed the pen had an average of 5.0 strokes (95\% CI: [4.3, 5.6]), while the panel had an average of 5.3 strokes (95\% CI: [4.2,	6.4]). However, the magnitude (1/3 of a stroke) is minimal in practical terms. No significant interaction effect were observed $F(2, 259) = 1.62, p = .200, \eta^2_p = .01$.

When targeting objects, there was a significant main effect of \f{technique} ($F(1, 259) = 328.82; p < .0001; \eta^2_p = .56$). Overall, panel required significantly more strokes ($M$ = 8.7, 95\% CI: [7.7, 9.8]) than viaCross ($M$ = 3.5, 95\% CI: [3.0, 4.0]), a difference of over five strokes on average. There was a significant \f{technique} $\times$ \f{difficulty} interaction $(F(2, 259) = 1.62, p = .020, \eta^2_p = .01)$. Panel required more strokes than viaCross, with the gap widening at higher difficulty: 3.7 vs. 2.9 (1 attribute; $p = .063$), 10.2 vs. 3.6 (2 attributes;  $p < .001$), and 12.3 vs. 3.9 (3 attributes;  $p < .001$).

When refining attributes, there was a significant main effect of \f{technique} ($F(1, 259) = 67.78; p < .0001; \eta^2_p = .21$). Panel required significantly fewer strokes ($M$ = 5.5, 95\% CI: [4.7, 6.5]) than viaCross ($M$ = 7.1, 95\% CI: [6.5, 7.8]). No significant interaction effect were observed $F(2, 259) = 1.00, p = .367, \eta^2_p = .01$.

When refining objects, panel required significantly ($F(1, 259) = 381.74, p < .001, \eta^2_p = .596$) more strokes than viaCross overall ($M$ = 10.1 vs. 3.8, 95\% CI [9.1, 11.1] vs. [3.5, 4.2]), and this difference increased with \f{difficulty} ($F(2, 259) = 32.44, p < .001, \eta^2_p = .200$): 5.0 vs. 2.4 (1 attribute; $p < 0.001$), 10.3 vs. 3.7 (2 attributes; $p<0.0001$); 15.0 vs. 5.4 (3 attributes; $p<0.001$).

\paragraph{Summary} viaCross offers comparable or superior efficiency to panel in most tasks, with the largest benefits emerging for object-level selection and refinement as task difficulty increases. For attribute-level tasks, differences between techniques are minimal.


\subsection{Task Duration}
When targeting attributes, there was a significant main effect of \f{technique} ($F(1, 259) = 95.40; p < .0001; \eta^2_p = .27$), where viaCross took on average 8.10s longer than the panel.

When targeting objects, there was a significant main effect of \f{technique} ($F(1, 259) = 27.22; p < .0001; \eta^2_p = .10$) and an interaction effect of \f{technique} $\times$ \f{difficulty} ($F(2, 259) = 13.74; p < .0001; \eta^2_p = .10$). There was no significant difference in completion time between panel ($M = 8.28$s, 95\% CI: [7.16s, 9.72s]) and viaCross ($M = 10.21$s, 95\% CI: [8.43s, 12.24s]) for one attribute ($p = .100$). However, at higher difficulty, panel was significantly slower: for two attributes, panel averaged 25.21 s (95\% CI: [20.44s, 30.90s]) versus viaCross of 22.44s (95\% CI: [15.06s, 31.76s]; $p < .001$), and for 3 attributes, panel averaged 32.35s (95\% CI [27.13s, 38.34s]) versus viaCross of 22.67s (95\% CI: [16.78s, 30.44s]; $p < .001$).

When refining attributes, there was a significant main effect of \f{technique} ($F(1, 259) = 384.98; p < .0001; \eta^2_p = .60$), where viaCross took on average 21.5s longer than the panel.

When refining objects, there was a significant main effect of \f{technique} ($F(1, 259) = 56.74; p < .0001; \eta^2_p = .18$) and an interaction effect of \f{technique} $\times$ \f{difficulty} ($F(2, 259) = 29.61; p < .0001; \eta^2_p = .19$). Mean duration increased with \f{technique} and \f{difficulty}. For the panel, the mean duration was 10.26s (95\% CI: [8.76s, 12.00s]) for one attribute, 24.76s (95\% CI: [21.55s, 28.37s]) for two attributes, and 41.22s (95\% CI: [34.93s, 48.38s]) for 3 attributes. For pen, mean duration was 13.38 (95\% CI: [10.73s, 16.30s]), 16.76s (95\% CI [13.72s, 20.30s]), and 22.56 (95\% CI [18.85s, 27.09s]), respectively. Duration clearly increased with \f{difficulty}, and panel tended to show higher duration than viaCross, particularly at higher \f{difficulty} levels.

\paragraph{Summary} Overall, viaCross is slower than the panel for attribute-driven tasks, but it scales better for object-level selection and refinement. At low difficulty, the two techniques perform similarly, yet as complexity increases, the panel slows dramatically while viaCross remains efficient. This highlights a clear trade-off: panels perform well with attribute filters, while viaCross is better for complex, object-centric selections and is comparable only in creating attribute selections.

\subsection{Accuracy}
A GEE logistic regression was conducted to analyze accuracy. The \f{difficulty} significantly affected accuracy (1 vs. 2: $b = 1.17, p = 0.012$; 1 vs. 3: $b = 1.24, p = 0.014$), indicating lower accuracy for higher \f{difficulty} (panel: 1 = 92\%, 2 = 85\%, 3 = 85\%; viaCross: 1 = 88\%, 2 = 83\%, 3 = 83\%). The main effect of \f{technique} was not significant ($b = 0.94, p = 0.093$). Interactions between  \f{technique} and \f{difficulty} were not significant ($p = 0.23$), suggesting that the effect of \f{difficulty} was similar across techniques.

\subsection{Participants' Comments}

Participants noted that viaCross required a short learning period but became easier and more enjoyable with practice. Users described it as \textit{``attribute selection was slightly challenging at first, with more practice it started to get easier'' - P18} and \textit{``a little difficult to understand in the beginning, but once I got the hang of it, it was easy to use'' - P4}. In contrast, while the panel was immediately intuitive for many, \textit{``easy and not challenging'' - P14} and \textit{``very quick and easy to use'' - P13}, some found it cumbersome for complex selections, reporting that it was \textit{``a little more frustrating when trying to find multiple specific items'' - P1} and \textit{``cumbersome when [selecting] for shape, size, and colour for multiple items at once.'' - P2}

Users highlighted the widgets as both engaging and effective. They described them as \textit{``really effective compared to the attribute selection'' - P13} and \textit{``interesting to use'' - P10}. The ability to filter by multiple attributes was noted as \textit{``pretty useful'' - P7} and \textit{``very precise and didn’t involve the tediousness of deselecting unrelated objects one by one'' - P17}, which suggests it helped users gain precise control over selections. Some minor challenges were noted as \textit{``[reopening] the widget menu was confusing'' - P16}, but participants emphasized that \textit{``with more practice [viaCross] is way easier to do all this stuff'' - P6}, reflecting increased efficiency and comfort with repeated interaction, but with a short learning curve.

Overall, participants appreciated both techniques for different reasons. The panel offered immediate accessibility and straightforward task completion, while viaCross provided precision, flexibility, and engagement for complex tasks. Participants found viaCross especially suitable for detail-oriented use cases.

\section{Discussion}

Our findings highlight the complementary strengths of panels and viaCross. Panels offered
immediate efficiency in simple attribute filtering, while viaCross scaled better as tasks grew
more complex, particularly for object-centric selections. These trade-offs structure the
discussion below.

\paragraph{Creating Attribute vs. Object Selection Trade-offs}
As task difficulty increased—with more distractors and attribute overlap—the panel’s
performance degraded, while viaCross was less affected. The panel required up to four strokes to select a single attribute and was prone to misclicks and repeated deselection. Enlarging the panel might mitigate these issues, but only at the cost of valuable screen space. By contrast, viaCross, though slower for simple attribute selections, matched panel performance for
multi-attribute selections after practice and outperformed it in complex, object-centric tasks. These results suggest value in hybrid or adaptive interfaces: panels remain well suited for rapid, low-complexity filtering, while viaCross supports exploratory or object-based selection at higher complexity.

%The shift in strengths from panel to viaCross under increasing difficulty reflects our task design: by adding distractors and attribute overlap, we approximated real-world complexity. Results show that panel performance degrades as task complexity grows, whereas viaCross is less affected by increases in difficulty. The panel required up to four strokes to select a single attribute and was prone to misclicks and repeated deselection. Enlarging the panel might reduce these issues at the cost of screen space. viaCross, while slower for creating attribute selections, performed comparably on multi-attribute selections with practice and was faster and more efficient on object-centric tasks. These results suggest the value of hybrid or adaptive interfaces: panels for rapid, low-complexity filtering, and viaCross for exploratory or object-based selection at higher complexity.

\paragraph{Reducing Errors in Refining Multi-Attribute Tasks}

viaCross enabled precise, fine-grained control, but refinement tasks were slowed by the need to recreate widgets from scratch. Participants reported that this process became faster with practice, yet it still added effort compared to the panel. Preserving widgets across edits or complementing gestures with lightweight WIMP controls could reduce this burden by maintaining user effort and lowering cognitive load. A key design implication is that refinement mechanisms should support persistence—either through lasting widgets or hybrid pen/panel controls.

%viaCross supported more precise and controlled selections, which for complex selections could be time-consuming and prone to errors. A limitation discovered in viaCross was that refining selections involved starting from scratch, forcing users to recreate widgets. Participants reported that with practice, attribute-based widget creation became faster, reducing the duration for refinement tasks. Performance could be further improved by preserving widgets during refinement or by providing a static WIMP-style interface, avoiding repeated recreation. This was reflected in comparable performance for creating attribute selections but slower performance for refinement. Preserving user effort through persistent widgets or hybrid pen/WIMP controls can reduce errors and cognitive load. These insights suggest that refinement mechanisms should preserve user effort, either by allowing widgets to persist across edits or by blending pen gestures with lightweight WIMP-style controls.

\begin{figure*}[ht]
    \centering
    \subfloat[]{\includegraphics[width=0.33\textwidth]{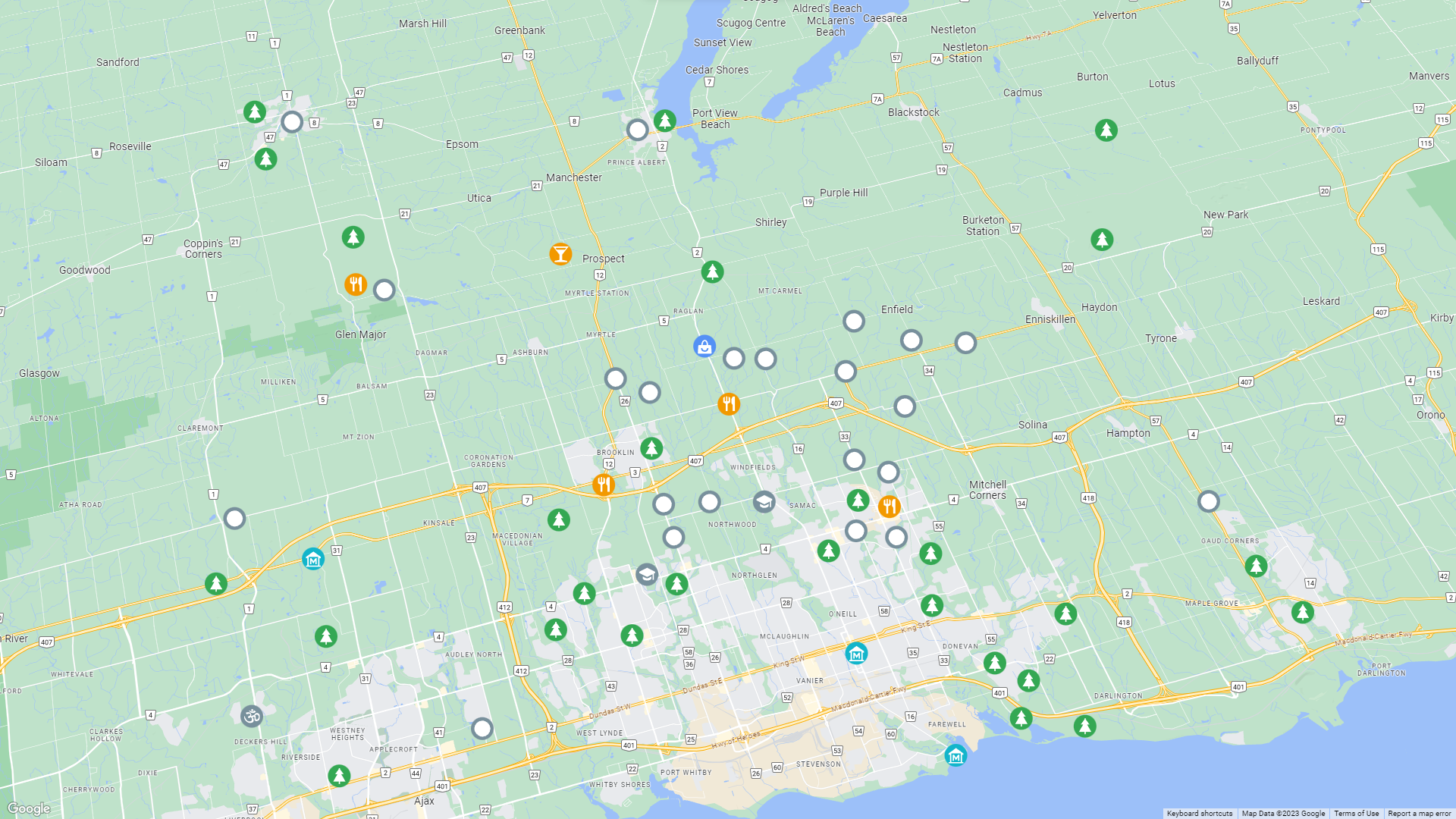}\label{fig:googleMaps1}}
    \hfill
    \subfloat[]{\includegraphics[width=0.33\textwidth]{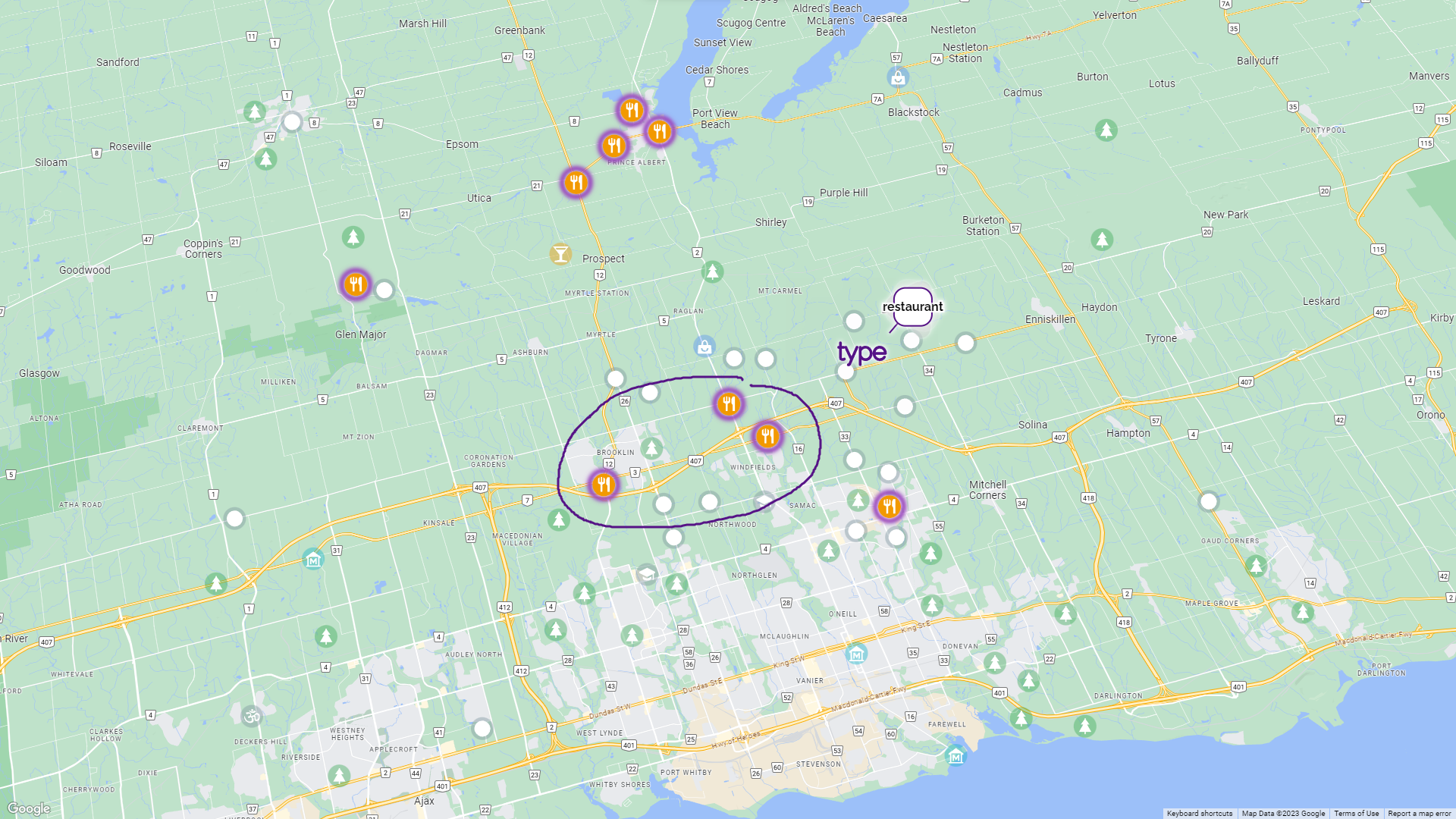}\label{fig:googleMaps2}}
    \hfill
    \subfloat[]{{\includegraphics[width=0.33\textwidth]{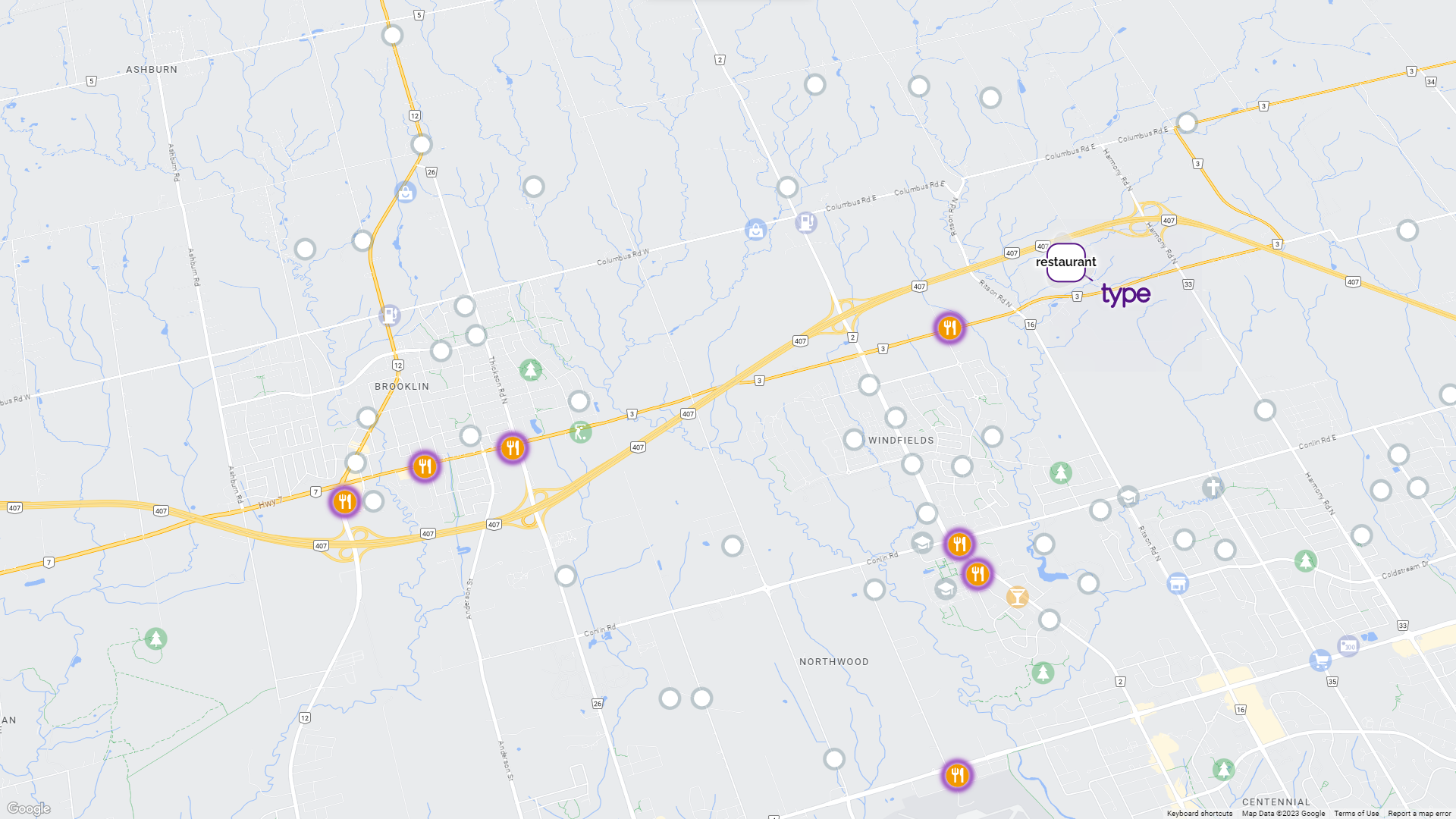}\label{fig:googleMaps3}}}
    \caption{Prototype implementation of viaCross within Google Maps.
\protect\subref{fig:googleMaps1} Initial map layout showing a mix of
locations. \protect\subref{fig:googleMaps2} Attribute-similarity selection highlights restaurants, bringing relevant markers to the foreground.
\protect\subref{fig:googleMaps3} Zooming into a circled region reveals additional restaurants and markers. This prototype demonstrates
the feasibility of applying viaCross principles to real-world map exploration, supporting filtering and
refinement without repeated menu navigation.
}
    % \caption{\protect\subref{fig:googleMaps1} Starting layout of Google Maps where various locations are displayed. \protect\subref{fig:googleMaps2} Using attribute similarity, Google Maps will search for restaurants in the area and will put restaurants at the forefront. Note that some restaurants were revealed that were not from~\nameref{fig:googleMaps1}. Circling a region will zoom into the region, revealing more markers \protect\subref{fig:googleMaps3}. The zoomed circled region from ~\nameref{fig:googleMaps2} with the same filter. Again, it will reveal more restaurants and other locations of interest.}
    \Description{Three subfigures, labelled (a) through (c), show stages of interaction within a map interface. (a) An initial Google Maps view displays a broad geographic region populated with numerous colour-coded location markers of varying types. (b) A purple border visually emphasizes a subset of markers, while non-relevant markers are visually de-emphasized. The highlighted set corresponds to restaurant markers. (c) A zoomed-in view of a circled region from (b) reveals additional markers of the highlighted type within a smaller geographic area.}
    \label{fig:googleMaps}
\end{figure*}
\paragraph{Learning Curve}
Participants described viaCross as initially challenging but quickly becoming easier and more helpful with practice. This short learning period may explain why attribute selections took slightly longer despite requiring a similar number of strokes as the panel. While the panel was immediately intuitive, participants found it increasingly tedious for complex tasks that required extra strokes and deselections. In contrast, viaCross became more efficient with
familiarity, enabling smoother object selections and refinements. Ultimately, participants were able to determine which selection type to use, and cross objects and attributes of interest with similar accuracy as the panel. A design implication is that gesture-based systems could benefit from onboarding aids or progressive disclosure: techniques that allow novices to learn basic operations quickly while unlocking expert efficiency as complexity increases.

These trade-offs are not limited to controlled tasks. They also appear in real-world domains, as we illustrate with a prototype implementation of viaCross for map exploration
(\autoref{sec:usecases}).

\section{Use Case: Map Exploration}

\label{sec:usecases}

To demonstrate feasibility beyond controlled layouts, we implemented viaCross in a Google Maps
prototype. In map-based tasks, users often need to select cities or regions based on attributes
such as population, category, or location. With viaCross, a single stroke specifies both the
selection type and attribute constraints, while attribute widgets enable in-place refinement
without requiring navigation back to a panel. For example, a user can cross through cities above
a population threshold, then adjust or refine the selection by editing the widget directly on
the map.

In our prototype, viaCross supports stroke-based searche and selection in Google Maps. When searching for
restaurants with ratings above 3/5, standard interactions require repeated queries and manual
filtering, while displaying all restaurants at once leads to visual clutter. With viaCross, a
user can select one restaurant and then cross through others to define attribute similarity,
dynamically highlighting relevant results within the visible region. Attribute widgets allow
further refinement by adjusting thresholds or categories directly on the map. As the user zooms
in, additional restaurants appear and can be incorporated into the selection, supporting
iterative exploration without repeated menu navigation (\autoref{fig:googleMaps}).

This prototype illustrates how the principles of viaCross can be realized in a real-world
application, extending beyond controlled tasks. While we did not conduct a formal evaluation of
the Google Maps implementation, it serves as a proof of concept showing that crossing-based
multi-object and multi-attribute selection can be integrated into geospatial exploration. Future
work should examine performance and usability in applied domains such as mapping, scatterplots,
or immersive environments.

\section{Limitations \& Future Work}

Our study was conducted in a constrained environment with simple layouts and minimal object overlap. This choice allowed us to isolate the core mechanics of viaCross—single-stroke multi-object and attribute selection—without confounds from layout density, occlusion, or domain-specific tasks. Establishing performance baselines under controlled conditions is a standard first step in evaluating novel interaction techniques. Future work should extend to more realistic scenarios such as scatterplots, dense workspaces, or 3D environments (e.g., VR with controllers and tabletops), where factors like visual clutter, occlusion, and larger physical workspaces may further affect performance. Although addressing visual clutter was not our focus, future work could integrate clutter-reduction strategies such as Minimum Description Length (MDL)~\cite{mdl}. The map exploration use case illustrates how viaCross principles could extend to such domains, but systematic evaluations in these settings remain an important direction for future work.

In terms of refinement, participants were prompted to refine predefined selections rather than
correct errors arising in naturalistic workflows. This may have influenced performance, and
further work should investigate refinement in open-ended tasks where corrections occur
organically.

Hardware constraints also shaped performance. Our study used a small pen-enabled display with
all objects and panels within arm’s reach, as well as a pen with an eraser tip. Larger
displays, multi-monitor setups, or different pen/touch hardware may alter efficiency and
ergonomics, suggesting a need for broader device evaluations.

A further limitation is that our study was short-term. Although participants quickly learned
the basics of viaCross, efficiency gains may depend on extended practice. Longitudinal studies
are needed to examine how performance, accuracy, and preference evolve over time, and whether
viaCross can surpass panels as users gain expertise.

Finally, we focused solely on selection tasks, excluding moving, deleting, or editing object attributes. Prior work on Object Oriented Drawing~\cite{OODrawing} and Attribute Spaces~\cite{attributeSpaces} in VR demonstrate how these operations can extend interaction workflows. Integrating viaCross with these forms of manipulation is an important next step toward making selection gestures part of end-to-end interaction pipelines.
% viaCross complements it to allow for object and attribute selection --- which is an example of a broader applicability of our technique.

\section{Conclusion}
% This research presented a pen-based technique, viaCross, that can be used for multi-object selection in environments. We compared its performance to a traditional filter panel that made use of tapping. The user study showed that the panel is faster, but viaCross required fewer strokes when targeting objects and more strokes when targeting attributes. Furthermore, the participants mentioned that viaCross was faster once they were more familiar with it. Participants achieve some level of success in selecting multiple attributes or spatial constraints using only strokes without using any other input (ex. hands or keyboard). We hope the results of this work will encourage new gestures and the use of widgets for pen-based selection techniques, potentially even in desktops or VR, to improve overall performance.

This research introduced viaCross, a pen-based crossing technique for multi-object selection, supported by attribute widgets for in-place refinement. Through a controlled comparative study with a panel-based WIMP interface, our study mapped the trade-offs between techniques: panels were faster and more intuitive for simple attribute filtering, while viaCross scaled better for complex, object-centric selections. 
Our findings highlight design implications for hybrid and adaptive interfaces, mechanisms that preserve user effort during refinement, and onboarding aids to support learning curves. The map exploration use case illustrates the potential generalization of these principles beyond controlled tasks, to domains such as geospatial analysis, scatterplots, or immersive environments. At the same time, limitations such as short-term evaluation and constrained layouts point to the need for longitudinal and ecological studies. 

By revealing when panels excel and when gestures scale, our work contributes not only a new technique but also broader insights for designing selection workflows across pen, touch, visualization, and immersive contexts.

% This research presents viaCross, a pen-based technique for multi-object selection, compared to a traditional WIMP interface with a filter panel and tapping. The user study showed that viaCross performs better for complex object-level selection and refinement, requiring fewer strokes and less time as task difficulty increases. While panels were faster and more intuitive for simple attribute-based selections, they suffered from repeated deselection and occasional misclicks. viaCross achieved comparable performance for creating attribute selections but was slower during refinement due to the need to recreate widgets, suggesting that preserving attribute widgets or providing a post-creation WIMP-style interface could improve efficiency. Overall, participants successfully used crossing-based gestures for multi-object selection. We hope this work encourages new crossing-based gestures and widget use for pen-based selection, potentially improving performance in desktops or VR.

\bibliographystyle{ACM-Reference-Format}
\bibliography{sample-base}

% \appendix
% \section{Additional Results}\label{sec:appendix-results}
% \input{sections/A1-results}

\end{document}